# Free electron topological bound state induced by light beam with a twisted wavefront

Yiming Pan[1*], Ruoyu Yin[2], Yongcheng Ding[3], Huaiqiang Wang[4], Daniel Podolsky[5], Bin Zhang[1,6*]

1. School of Physical Science and Technology and Center for Transformative Science, ShanghaiTech University, Shanghai 200031, China
2. Department of Physics, Institute of Nanotechnology and Advanced Materials, Bar-Ilan University, Ramat-Gan 52900, Israel
3. Department of Physical Chemistry, University of the Basque Country UPV/EHU, Apartado 644, 48080 Bilbao, Spain
4. Center for Quantum Transport and Thermal Energy Science, School of Physics and Technology, Nanjing Normal University, Nanjing 210023, China
5. Department of Physics, Technion, Haifa 3200003, Israel
6. Department of Electrical Engineering Physical Electronics, Tel Aviv University, Ramat Aviv 6997801, Israel

**Abstract**

Recent advances in ultrafast electron emission, microscopy, and diffraction have demonstrated a remarkable ability to manipulate free electrons with quantum coherence using light beams. Here, we present a framework for exploring free electron quantum number in ultrafast electron-light interactions. We derive an explicit Jackiw-Rebbi solution for a low-energy free electron wavefunction subjected to a spatiotemporally twisted laser field, resulting in a flying topologically protected bound state with a quantum number of e/2 – termed a "half-electron". This flying bound state is dispersion-free due to its topological nature. We demonstrate the topological confinement and pair generation mechanism of half-electrons in free space, expanding their domain beyond the topological states typically found in solids and photonics. This advancement enhances our understanding of emulating exotic quantum and topological effects with low-energy free electrons.

Modern physics allows fractional charges. There are two primary approaches to explore such exotic existence. One method examines the topological structure of the massive Dirac equation, initially demonstrated by Jackiw-Rebbi solutions in quantum field theory[1]. The other delves into many-body interactions in condensed matter systems, exemplified by Laughlin's wave function in Fractional Quantum Hall Effect (FQHE)[2]. Both realizations require specific materials such as polyacetylene[3], semiconductors[4], and artificial disclinations[5–8]. Traditionally, electrons were thought to carry only integer electric charges. However, in 1976, Jackiw and Rebbi challenged this conventional notion by proposing zero-energy bound states arising from the Dirac equation with a kink-profiled mass term. They argued that these zero-energy modes carry fractional charges, specifically e/2[1], due to their topological nature. This prediction was confirmed by Su, Schrieffer and Heeger's discovery in Polyacetylene in 1979[3]. Further evidence came from the FQHE[9], where a bunch of composite quasiparticles exhibited fractional charges.

Recently, numerous proposals and experiments are actively exploring fractional charges as topological bound states in quantum materials[10–12], and topological photonics with disclinations[13,14]. However, the detection of topological charges in materials and devices remains indirect, relying on methods such as local denisty of states, shot noise measurement, edge state current, and Aharonov-Bohm oscillations[4,15–19]. Direct utilization of these topological charges is still far from being feasible. These intriguing developments raise a compelling question: Is it possible to construct material-free topological states with free electrons, and if so, can we manipulate the flying bound state?

It is noteworthy that our proposal of a free electron bound state aligns well with the advancing frontier of attosecond physics. Recent breakthroughs in ultrafast free electrons have revolutionized electron microscopy[20], diffraction[21,22], and even free electron radiation and lasing[23,24]. Electron beams have achieved extremely high temporal resolution on the femtosecond and attosecond scales[25–27]. However, a significant challenge remains for low-energy electrons: the nonrelativistic dispersion and propagation chirping of ultrashort electron pulses, even in a vacuum. For instance, slow electrons operating at energy of 100 eV are not immune to dispersion. Understanding and controlling the dispersion effect is crucial for harnessing and characterizing attosecond electron pulses in innovative experiments, as recently demonstrated by [12,28–30]. The electron wavefunction engineering holds enormous potential, offering deeper insights into the quantum realm and paving the way for groundbreaking technological advances[12,31–33]. Considering the possibility of generating and manipulating free electron topological charges, our focus now shifts to recent developments in low-energy electron wavefunction and quantum light interaction. These advancements have unveiled fascinating dispersion effects, such as free electron Rabi oscillations and self-trapping in energy space[34–38].

Here, we present a pioneering strategy that leverages the dispersion of wide spectral sidebands to construct a pulsed "two-level" electron wavefunction, involving only two photon-induced sidebands as a spinor (see Fig. 1b)[34]. This Dirac-type interaction arises from the coupling of a slow electron pulse with a monochromatic light field with a spatiotemporally twisted wavefront, mimicking the kink profile of the mass term in the celebrated Jackiw-Rebbi model[1]. The resulting light-induced kink inherits a quantum number, yielding a topological fractional charge of e/2, which we refer to as a "half-electron." This flying topological bound state propagates in a dispersion-free manner, a striking feature for manipulating low-energy slow electron pulses using a light beam. Additionally, we propose a time-of-flight streaking experiment to demonstrate the topological confinement and pair generation of these light-induced half-electrons.

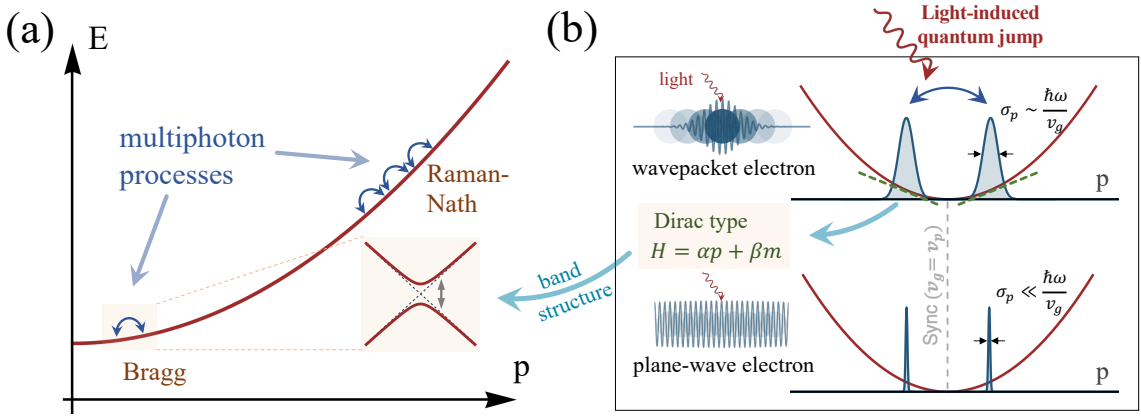

**Figure 1:** Construction of low-energy free-electron Dirac equation by light. (a) Multiphoton processes, represented on the energy dispersion, and for the slow (fast) electron interaction, Bragg (Raman-Nath) diffraction occur. In Bragg regime (orange region), where only two sidebands involved, we construct the effective Dirac-type equation of the "two-level" free electron. Specifically, as seen in (b), the energy spread of the point-like electron plays the momentum term of the Dirac equation, while the light-electron coupling strength plays the Dirac mass term.

**Constructing slow electron Dirac equation by light.** The nonrelativistic low-energy electron Hamiltonian in the presence of electromagnetic field is given by $H = \frac{(\mathbf{p}-e\mathbf{A})^2}{2m}$. We assume the longitudinal vector potential and the electric field are modulated by a grating, given by $A = \frac{E_0}{\omega}\cos[\omega t - qz - \theta(z,t)]$, and $E = -\frac{\partial A}{\partial t} = E_0 \sin[\omega t - qz - \theta(z,t)]$, respectively, where $\omega$ is the frequency, $q$ is the wavevector, and $\theta(z,t)$ is the initial phase. A spatiotemporally-varying kink profile would be imprinted on this phase term, which we will discuss it later. Note that the vector potential on the grating, as well as the electric field, is along the traveling direction of the electron, $z$ direction.

Such modulation of slow electrons by electromagnetic fields, or essentially electron-photon interactions, described by the time-dependent Schrodinger equation (TDSE), can be classified through the Raman-Nath and Bragg regimes, each distinguished by distinct phase matching conditions and coupled-mode equations[34,35,38]. In the Raman-Nath regime, characterized by weak-field coupling, the phase matching condition ensures efficient interaction when the electron's wavevector aligns with that of incoming photons, leading to multiphoton scattering events and phase accumulation, as shown in Fig. 1a. Conversely, the Bragg regime entails strong-field interactions, with a modified phase matching condition accounting for the periodicity of the synthetic lattice[34]. In this regime, electrons undergo Bragg diffraction in synthetic dimension, with their momentum quantized by light, enabling precise control and manipulation. The coupled-mode equations, which are vital for our discussions in the Bragg regime under investigation in this Letter, are simplified in the momentum representation, and in the co-moving frame ($z \to z - \omega t/q$) as

$$i\hbar \, \partial_t |\delta k - q/2\rangle = \left[\frac{\hbar^2 \delta k^2}{2m} - \left(\frac{\hbar^2 q}{2m}\right)\delta k\right]\left|\delta k - \frac{q}{2}\right\rangle - \kappa^* \left|\delta k + \frac{q}{2}\right\rangle,$$
$$i\hbar \, \partial_t |\delta k + q/2\rangle = \left[\frac{\hbar^2 \delta k^2}{2m} + \left(\frac{\hbar^2 q}{2m}\right)\delta k\right]\left|\delta k + \frac{q}{2}\right\rangle - \kappa \left|\delta k - \frac{q}{2}\right\rangle. \quad (1)$$

Here the coupling coefficient $\kappa = \frac{eE_0 \hbar k}{2m\omega} e^{i\theta(z,t)}$, depending on the amplitude strength $E_0$ and the spatiotemporally-varying phase profile $\theta(z,t)$. $\delta k$ is a shifted momentum corresponding to the momentum spread of the electron wavepacket and the original momentum is $k = k_0 + \delta k$ since k is a continuous number. For convenience, we shift $\delta k \to \delta k \pm q/2$ for denoting two sidebands with q being the photon induced quantum recoil. In the condition $\frac{\hbar^2 (q/2)^2}{2m} \gg |\kappa|$, namely using the approach of adiabatic elimination[39] to truncate the coupled modes into the case involving two sidebands[34,35]. See Methods for more details of the truncation. Defining the wavefunction as a spinor $|\tilde{\psi}\rangle = \alpha \left|\delta k - \frac{q}{2}\right\rangle + \beta \left|\delta k + \frac{q}{2}\right\rangle = [\alpha, \beta]^T$, Eq. 1 can be rewritten as

$$i\hbar\partial_t|\tilde{\psi}\rangle = \frac{\hbar^2 \delta k^2}{2m}|\tilde{\psi}\rangle - \frac{\hbar^2 q}{2m}\delta k \cdot \sigma_z|\tilde{\psi}\rangle - |\kappa|e^{-i\theta\sigma_z}\sigma_x|\tilde{\psi}\rangle \qquad (2)$$

Correspondingly, the effective Dirac Hamiltonian is constructed as

$$H_D = \frac{\hbar^2 \delta k^2}{2m}\sigma_0 - \frac{\hbar^2 q}{2m}\delta k \cdot \sigma_z - |\kappa|e^{-i\theta\sigma_z}\sigma_x \qquad (3)$$

where $\sigma_0$ is an identity matrix, $\sigma_z$ and $\sigma_x$ are the Pauli-$z$ and Pauli-$x$ matrices. Notice that the first term is a dispersion term, it is negligible in the approximation $\delta k < q$. For the case of a constant phase $\theta$, we obtain the Dirac dispersion $E_\pm = \pm\sqrt{\left(\frac{\hbar^2 q}{2m}\delta k\right)^2 + |\kappa|^2}$, see the inset of Fig. 1a.

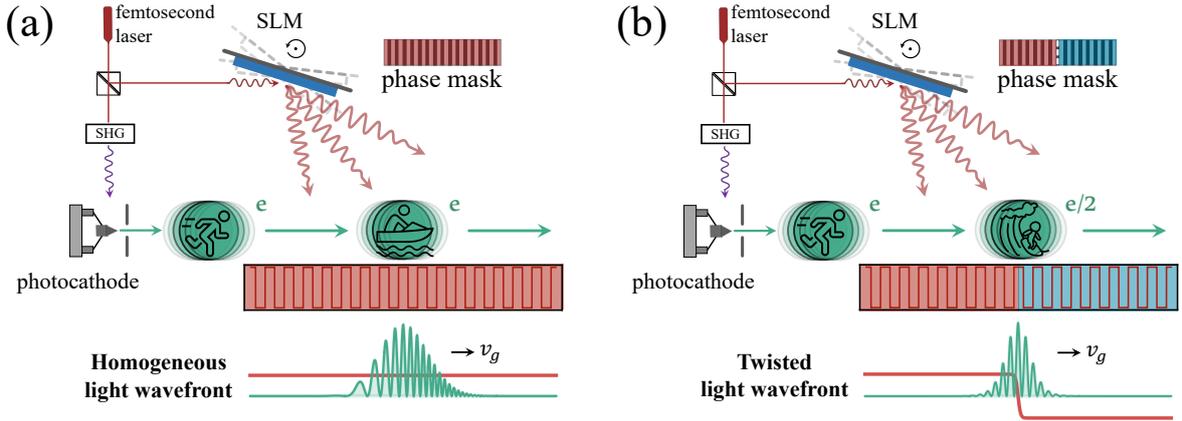

**Figure 2:** The schematic setup of the interaction of (a) one-electron (with electric charge of e) or (b) half-electron (carrying topological charge of e/2) with a homogeneous or spatiotemporally twisted laser beam. We design a twisted wavefront of the light beam to manifest a flying "kink" structure by utilizing a spatial light modulator (SLM).

**Free electron Jackiw-Rebbi solutions.** We consider a simple Dirac mass term with a kink structure, the wavefront profile of the laser beam as shown in Fig. 2b,

$$\kappa(z,t) = \kappa\left(z - \frac{\omega t}{q}\right) = |\kappa|e^{-i\theta(z-\omega t/q)\sigma_z} = \begin{cases} +|\kappa|, & \text{for } z - \omega t/q = -\infty \\ -|\kappa|, & \text{for } z - \omega t/q = +\infty \end{cases} \qquad (4)$$

allows to calculate the zero-mode solution ($i\hbar\partial_t\psi_0 = 0$)

$$\left(-\frac{\hbar^2 q}{2m}\delta k \cdot \sigma_z - |\kappa|e^{-i\theta\sigma_z}\sigma_x\right)\psi_0 = 0 \tag{5}$$

We neglect the first quadratic term in Eq. 3 and recall the spatial representation,

$$\left[-i\frac{\hbar^2 q}{2m}\sigma_z \partial_z + \kappa(z)\sigma_x\right]\psi_0 = 0 \tag{6}$$

Multiplying both sides of Eq.6 from the left with $\sigma_z$ yields $\left[-i\frac{\hbar^2 q}{2m}\partial_z + i\kappa(z)\sigma_y\right]\psi_0 = 0$. Hence, we write the zero-mode solution as $\psi_0(z) = \chi_\pm^{(y)}\varphi_\pm(z) = \frac{1}{\sqrt{2}}\begin{bmatrix}1\\\pm i\end{bmatrix}\varphi_\pm(z)$, with $\varphi_\pm(z)$ determined by $\frac{\hbar^2 q}{2m}\partial_z\varphi_\pm(z) = \pm\kappa(z)\varphi_\pm(z)$. Therefore, we obtain an explicit Jackiw-Rebbi solution of free electron, as given by $\psi_0^{(\pm)} = \frac{1}{\sqrt{2}}\begin{bmatrix}1\\\pm i\end{bmatrix}\text{Exp}\left[\pm\frac{2m}{\hbar^2 q}\int_0^\xi \kappa(\xi')d\xi'\right]$, where $\xi = z - \omega t/q$. Since one of the solutions diverges exponentially ($\psi_0^{(-)}$) due to kink profile (Eq. 4), we abandon this unphysical solution that cannot be normalized. Only one solution with a special chirality ($\psi_0^{(+)}$) remains. Roughly speaking, this chiral zero mode would lead to half-electron (e/2), which will be expound later. Return to the original representation of the time-dependent Schrodinger equation, we express the free electron bound state explicitly as

$$\psi_0(z,t) = \left(\frac{e^{-i(qz-\omega t)/2} + i\, e^{i(qz+\omega t)/2}}{\sqrt{2}}\right) e^{ik_0 z - i\left(\frac{\hbar q^2}{8m}\right)t}\, e^{\frac{2m}{\hbar^2 q}\int_0^{z-\omega t/q}|\kappa(\xi')|d\xi'} \tag{7}$$

To achieve and implement the half-electron solution, we employ a laser beam with a phase profile modulated by a Spatial Light Modulator (SLM). As illustrated in Fig. 2, the femtosecond laser beam is split into two paths. One path stimulates the photocathode to emit electrons, following a Second Harmonic Generation (SHG) that generates UV light. The other beam is modulated by the SLM to form a homogeneous or twisted wavefront of the coupling laser field. An optical grating (represented by the colored blocks with squared wave inside) is utilized to guarantee the phase matching condition during interactions. As a calibration, Fig. 2(a) shows that an electron remains as a carrier of one unit of charge under the homogeneous wavefront, exihibiting dispersive behavior. While if we use an electrically-controlled rotating SLM to twist the beam wavefront, a solitary wave carrying topological charge of e/2 (from the viewpoint of topological insulator[40]) "surfs" on the kink structure of the twisted wavefront. We choose a comoving kink profile as $\kappa(\xi) = -|\kappa|\tanh(\xi)$, and the profile of the solitary solution is proportional to $\text{sech}(\xi)$, exhibiting light-induced non-spreading feature (compared to Eq. A1 in Methods).

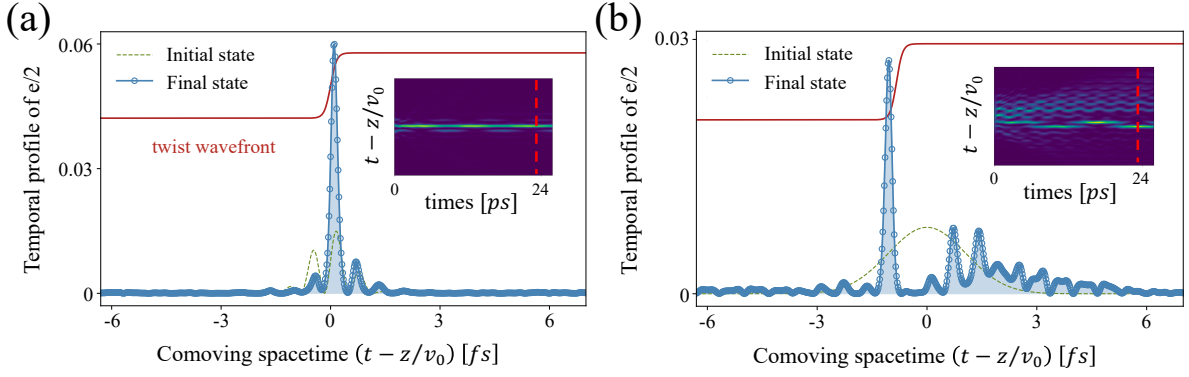

**Figure 3:** TDSE simulation of the temporal profile of a "half-electron (e/2)" induced by a twisted light wavefront. (a) The electron wavefunction is initially prepared as an explicit Jackiw-Rebbi solution (Eq. 7). The half-electron temporal profile is pinned at the kink structure (red curve) with a duration of less than 1 fs, comparable to the kink's size. Inset: Dispersion-free behavior of the half-electron evolving from 0 to 25 ps, showing topological protection. (b) The input electron is a Gaussian wavepacket, slightly offset from the kink center. The electron splits: one part is trapped by the kink, while the rest scatters into the bulk. Inset: Evolution pattern demonstrates the stability of the protected spectrum of half-electron, apart from the scattered part.

**TDSE simulations.** Figure 3 gives a direct simulation of the half-electron, using the time-dependent Schrodinger equation in the presence of the twisted light wavefront. An electron wavepacket, profiled as a Gaussian in the momentum space, featuring a center energy of 100 eV ($\beta = v_0/c = 0.02$), is manipulated such that the dispersion from the center energy is larger than the laser photon energy ($\lambda = 200$ nm and $\hbar\omega = 6.2$ eV). Subsequently, the electron traverses a grating structure designed to synchronize the light-induced near field with the electron, specifically $\omega/q = v_0$. Crucially, the center of the electron wavepacket undergoes fine-tuning to align with the center of the twisting structure of the laser wavefront. For the latter we chose a specific twisted phase profile $\theta(z,t) = \frac{\pi}{2}\tanh\left(\frac{z-v_0 t}{\eta}\right)$ to achieving the flying kink (with the parameter $\eta = 0.001$). The near field interacts in the grating structure with the electron, through the minimal coupling $A \cdot p$ in the time-dependent Hamiltonian. The evolving electron wavefunction in space and with time is recorded and plotted in the insets of Fig. 3. Notably, when comoving with the electron's pulse center, the soliton profile becomes evident (Eq. 7), and the dispersion-free property, a consequence of topological protection, is illustrated in the inset of Fig. 3. When the electron center is slightly offset to the twist (with a distance of 0.8 fs), the electron splitted into a kink-trapped part and a light-scattered part. The flying bound state imposes self-constraints over time. The details of TDSE simulation see Supplementary Information.

**Topological nature of half-electrons.** As we will prove from quantum field theory, this is a twisted light beam that leads to electron state possessing a fractional charge, indicating topological protection by the twisted wavefront. We can tune the phase of the light wavefront $\theta(z,t)$ to show the fractional charge of the zero-energy solution (Eq.7). By rewriting the effective Dirac Hamiltonian $H_D$ in the field-quantization formulation, we obtain

$$\widetilde{H}_D = \int dz\, \Psi^\dagger(z) \left( \frac{\hbar^2 \delta k^2}{2m} \mathbb{1}_2 - \frac{\hbar^2 q}{2m} \delta k \cdot \sigma_z - |\kappa| e^{-i\theta \sigma_z} \sigma_x \right) \Psi(z) \qquad (8)$$

where the electron field operator $\Psi(z)$ and $\Psi^\dagger(z)$ are the annihilation and creation operators in the second quantization picture. We directly apply the Goldstone-Wilczek formula[41], and then obtain the 2-current $J^\mu = (\rho, j_z) = \frac{e}{2\pi} \epsilon^{\mu\nu} \partial_\nu \theta$. Hence, the transferred charge $\Delta Q$ for a spatiotemporally varying phase, $\theta(z,t)$, is determined by

$$\Delta Q = -\frac{e}{2\pi} \int dz\, \partial_z \theta(z) = -\frac{e}{2} \qquad (9)$$

This indicates that tuning the phase of the light beam, which interacts with low-energy electrons under synchronization condition, will drive the anomalous current and generate a topological fractional charge. Specifically, the local domain wall carries a half-charge of $\pm e/2$, with the sign determined by the phase profile difference between boundary values, $\theta(+\infty) - \theta(-\infty)$. Splitting electrons into half-electrons is a pivotal concept, revolving around the pair generation and annihilation of half-electrons, with each kink contributing an increment of $e/2$ to the overall charge, resulting in one electron. It is worth mentioning that in the celebrated SSH model, the two half-charged bound states originating from anomaly are located at the boundaries in an inversion-symmetric manner[42]. Figure 4 demonstrates the evolution pattern of a half-electron pair that was constructed by a double-twisted wavefront of the laser field.

Note that the light-induced half-electron goes beyond a simple combination of "a kink and a fermion". It arises from the interaction between electromagnetic fields and free electrons, akin to the topological soliton in polyacetylene when Bloch electrons resonantly interact with phonons. The half-electron, a flying topological bound state, combines aspects of bosonic and Dirac-type fermionic fields. To avoid complications in interpretating how light splits one electron into two half-electrons, we view the half-electron purely as a topological bound state possessing half the degrees of freedom of an "normal" electric charge. This is similar to the fractional excitations that hosts a fraction of the total number of states demonstrated in topological photonic materials[7,8]. This interpretation clarifies the nature of half-electrons as a novel light-induced topological phenomenon.

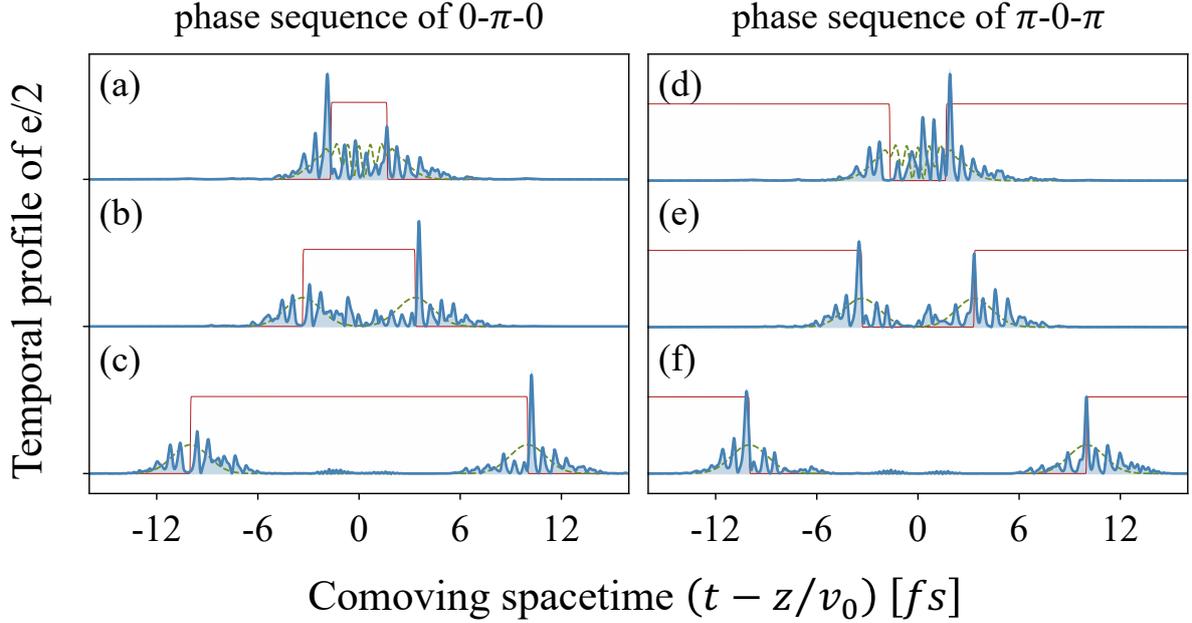

**Figure 4:** Demonstration of free-electron kink-antikink pair generations. The left column shows the phase twist profile $\theta(z,t)$ transitioning from 0 to $\pi$ and back to 0, while the right column shows $\theta(z,t)$ transitioning from $\pi$ to 0 and back to $\pi$. (a, d) have an interval of 3.3 $fs$, resulting in overlapping electron wavepackets and the beating pattern term. The beating allows the hybridization of two half-electrons. (b, e) have a 6.6 $fs$ interval, with slightly overlapped temporal profile. (c, f) have an 20 $fs$ interval, showing well-localized temporal profile around the kink and antikink structures with evanescent tails.

**Further discussions.** Due to the topological nature of the half-electron, its confined temporal profile depends on the specific spatiotemporal configuration of the twisted light wavefront (Eq. 4 and 7). This light-induced temporal profile can be measured using time-of-flight detector or a longitudinal electron interferometer by assessing its coherent length. Additionally, their dispersion-free propagation of the flying bound states, unlike that of slow electrons, allows for experimental calibration of signals indicating strong confinement (or dispersion) by switching the light field on (or off).

This half-electron leverages a 'clean quantumness' to advance quantum simulations using low energy electron wavepackets. Free electrons, fundamental since the inception of quantum mechanics, exhibit manipulatable coherence and interact readily with electromagnetic waves. This allows fine-tuning of their dispersion and coupling, as demonstrated by the creation of flying bound states (Figs. 3 and 4). Echoing previous work[43], our results show that manipulating quantum light interacting with electrons can profoundly shape the latter. Recent advances have enabled information transfer between light and electrons within a photon-electron-entangled system[44]. Our proposal, particularly constructing an effective Dirac equation and free electron topological bound state through a twisted light beam, signifies a novel method for quantum simulations[45–49]. This approach, using femtosecond laser to control low energy electrons, could revolutionize our understanding and manipulation of exotic quantum systems.

**Conclusions and outlook.** In brief, we have constructed the Jackiw-Rebbi solution for low energy free electrons using a spatiotemporally-twisted laser field, discovering a topological fractional charge of e/2, termed as a "half-electron". This light-induced quantum state exhibits a topological nature. Our work elucidates the emergence of flying topological bound state in vacuum rather in material. Our free-electron construction has profound implications, extending the domain of exotic topological excitations beyond conventional materials realms. We hope that such entities could serve as novel sources and applications for ultrafast electron diffraction, microscopy and spectroscopy.

Additionally, our work suggests a paradigm shift in studying the topological physics of free electrons, utilizing a material-free approach. By employing multiphoton process to mimic lattice structures such as photon-induced near-field electron microscopy[20], we open avenues to explore phenomena traditionally reserved for condensed matter physics within ultrafast free electron systems. This framework allows free electrons to emulate bound (or quasi-free) electrons in solids, trapped by laser fields rather than atomic potentials, achievable through all-optical control. Consequently, our construction holds promise for an emerging direction of "free-electron condensed matter physics", complementing recent studies on free-electron synthetic dimensions[35,36,38]. Further exploration of space-charge interactions within multi-electron beams[50,51] may reveal new possibilities for emulating fantastic physics such as strongly correlated systems.

**Acknowledgements**

We thank Hua Jiang, Dandan Hui for insightful discussions. Y.P. is supported by the National Natural Science Foundation of China (Grant No. 2023X0201-417).

The authors declare no competing financial interests.

Correspondence and requests for materials should be addressed to Y.P. (yiming.pan@shanghaitech.edu.cn) and B.Z. (zbphy28@gmail.com)

**Methods:**

**A. Dispersion and chirp of slow electron wavepacket** We can express a quantum electron wavefunction in term of a chirped Gaussian wavepacket.

$$\psi_0(z,t) = \frac{e^{i(p_0 z - E_0 t)/\hbar}}{(2\pi\sigma_z^2)^{1/4}\sqrt{1+i\xi t}} \exp\left(-\frac{(z-v_0 t)^2}{4\sigma_z^2(1+i\xi t)}\right) \quad (A1)$$

with the free-space chirp factor $\xi = \frac{\hbar}{2m^*\sigma_z^2}$ and the intrinsic wavepacket waist $\sigma_z = \frac{\hbar}{2\sigma_p} = \frac{\hbar v_0}{2\sigma_E} = v_0\sigma_t$ and the effective mass $m^* = \gamma^3 m$ with the Lorentz factor $\gamma$. Especially for low energy free electron ($\gamma \approx 1$), the seoncd-order dispersion is unavailable. The chirpig effect can be expressed in term of the electron width spreading, given by

$$\langle \sigma_z(t) \rangle = \sqrt{\sigma_z^2(1+\xi^2 t^2)} \quad (A2)$$

For our concern, we choose the kinetic energy of the electron $E_0 = 100\ eV$ and its group velocity is being $v_0 = \beta c$ with $\beta = 0.02$.

Slow electron experiences stronger dispersion. This is a problem in the development of low energy electron quantum optics. In Bragg regime, only two sidebands are involved, but the sideband has a spectral width. That width actually corresponds to the electron pulse duration with the Heisenberg uncertainty in the quantum limit. Especially, when the sideband width reaches the energy spacing between two sidebands, we understand that the Bragg regime is not suitable anymore. Indeed, we should consider the spectral spread of each sideband.

**B. Derivation of the coupled-mode equations Eq. (1)** Here we show the derivation of the coupled-mode equations Eq. (2). Starting from the Hamiltonian Eq. (1), to which we substitute the expression of $A = \frac{E_0}{\omega}\cos[\omega t - qz - \theta(z,t)]$, and neglecting the pondermotive term ($A^2$), we obtain $H = \frac{p^2}{2m} - \frac{eE_0}{m\omega}\cos[\omega t - qz - \theta(z,t)]\mathbf{p} - \frac{eE_0\hbar q}{2im\omega}\sin[\omega t - qz - \theta(z,t)]$. To deal with the problem in the comoving frame, we now define the operator

$$U(t) = e^{-i\frac{p\omega t}{\hbar q}} \quad (B1)$$

and it follows the identity $U(t)f(z) = e^{-\frac{\omega t}{q}\partial_z}f(z) = f(z - \frac{\omega t}{q})$. Assuming that the solution ψ to our Hamiltonian (1) has the transform $\psi = U(t)\phi = U\phi$, and multiplying the Schrödinger equation, $i\hbar \partial_t \psi = H\psi$, from the left by $U^\dagger = e^{i\frac{p\omega t}{\hbar q}}$, we obtain $i\hbar \partial_t \phi = (U^\dagger HU)\phi - i\hbar[U^\dagger \partial_t U(t)]\phi$, which indicates an effective Hamiltonian for the state φ,

$$H_{eff} = U^\dagger HU - i\hbar[U^\dagger \partial_t U(t)] \tag{B2}$$

Direct calculation of each term in the right-hand side yields

$$H_{eff} = \frac{p^2}{2m} - \frac{\omega}{q}\mathbf{p} - \frac{eE_0}{m\omega}\cos(qz+\theta)\mathbf{p} + \frac{eE_0\hbar q}{2im\omega}\sin(qz+\theta) \tag{B3}$$

Transformed into the momentum representation, using the basis of momentum eigenstate $|k\rangle$ ($p = \hbar k$), and the position operator in momentum space $z = i\hbar \partial_p$, we find

$$H_{eff}|k\rangle = \left[\frac{\hbar^2 k^2}{2m} - \frac{\omega}{q}(\hbar k)\right]|k\rangle$$

$$-e^{i\theta}\left(\frac{eE_0\hbar k}{2m\omega} + \frac{eE_0\hbar q}{4m\omega}\right)|k-q\rangle - e^{-i\theta}\left(\frac{eE_0\hbar k}{2m\omega} - \frac{eE_0\hbar q}{4m\omega}\right)|k+q\rangle \tag{B4}$$

Here we use $\cos(qz+\theta) = [e^{i(qz+\theta)} + e^{-i(qz+\theta)}]/2$, $\sin(qz+\theta) = [e^{i(qz+\theta)} - e^{-i(qz+\theta)}]/2i$. For our concern, $|k| \gg q$, we can neglect the detuning term $\frac{eE_0\hbar q}{4m\omega}$ in the couplings, and then obtain an effective coupled-mode equations,

$$i\hbar \partial_t |k\rangle = \left(\frac{\hbar^2 k^2}{2m} - \hbar\omega \frac{k}{q}\right)|k\rangle - \kappa|k-q\rangle - \kappa^*|k+q\rangle \tag{B5}$$

with $\kappa = \frac{eE_0\hbar k}{2m\omega}e^{i\theta(z,t)}$, depending on the amplitude strength $E_0$. Since the momentum k is a continuous number, we set $k = k_0 + \delta k$, and expand the on-site potential term of Eq. (6), then it becomes $\frac{\hbar^2(k_0+\delta k)^2}{2m} - \hbar\omega\frac{k_0+\delta k}{q} = \left(\frac{\hbar^2 k_0^2}{2m} - \hbar\omega\frac{k_0}{q}\right) + \left(\frac{\hbar k_0}{m} - \frac{\omega}{q}\right)\hbar\delta k + \frac{\hbar^2 \delta k^2}{2m}$. We apply the synchronization condition $\frac{\hbar k_0}{m} - \frac{\omega}{q} = 0$, corresponding to $v_g^{(e)} = v_p^{(ph)}$ (the group velocity of electrons equals to the phase velocity of light beam). This condition can locate the central momentum of the electron $p_0 = m\omega/q$, and the energy $E_0 = \frac{p_0^2}{2m} = \frac{m\omega^2}{2q^2}$ (in our setting

$E_0 \sim 100$ eV). Under the synchronization condition, the coupled-mode equations (B5) are simplified as

$$i\hbar\, \partial_t |\delta k\rangle = \left(\frac{\hbar^2 \delta k^2}{2m} - \frac{\hbar^2 k_0^2}{2m}\right)|\delta k\rangle - \kappa|\delta k - q\rangle - \kappa^*|\delta k + q\rangle \tag{B6}$$

from which one can obtain Eq. (2) in the main text, through shifting $\delta k \to \delta k \pm q/2$, and absorbing the constant on-site term $\frac{\hbar^2}{2m}\left(\frac{q^2}{4} - k_0^2\right)$ into the phase of the final solution.

# Supplementary information for
# Free electron topological bound state induced by light beam with a twisted wavefront

Yiming Pan[1*], Ruoyu Yin[2], Yongcheng Ding[3], Huaiqiang Wang[4], Daniel Podolsky[5], Bin Zhang[1,6*]

1. School of Physical Science and Technology and Center for Transformative Science, ShanghaiTech University, Shanghai 200031, China
2. Department of Physics, Institute of Nanotechnology and Advanced Materials, Bar-Ilan University, Ramat-Gan 52900, Israel
3. Department of Physical Chemistry, University of the Basque Country UPV/EHU, Apartado 644, 48080 Bilbao, Spain
4. Center for Quantum Transport and Thermal Energy Science, School of Physics and Technology, Nanjing Normal University, Nanjing 210023, China
5. Department of Physics, Technion, Haifa 3200003, Israel
6. Department of Electrical Engineering Physical Electronics, Tel Aviv University, Ramat Aviv 6997801, Israel

## 1. The full phase-space profiles of the half electron

In our simulations, we consider an electron with center energy 100 eV, corresponding to the group velocity of $v_0 = \beta c$ with the relative speed $\beta = 0.02$ and the Lorentz factor $\gamma = 1.001$. The electron is characterized by a Gaussian wave packet in the momentum space. To achieve synchronization between the phase velocity of light (of wavelength 200 nm) and the group velocity of the electron, we employ a nanograting with a period $\Lambda = 4$ nm.

Conducting a simulation based on the time-dependent Schrödinger equation, as elaborated below, the electron interacts with the light field over a time duration of 6 ps (divided into steps). By directly recording the amplitude of the wave function at each time step, we obtain the temporal profile of the half-electron's spatial distribution. Subsequently, performing a Fourier transform on the recorded wave function to recover the momentum representation, we derive the temporal profile of the half-electron's momentum distribution.

Below we show extended figures, as supplemental materials to Fig. 3 in the main text. We see the initial two sidebands in the momentum space, become smaller in amplitudes and merge with each other, as time elapses, leading to the appearance of more sidebands. And the spatial wave function, exhibits clearly solitary feature, which is manifested by the dispersion-free fact, as time goes by.

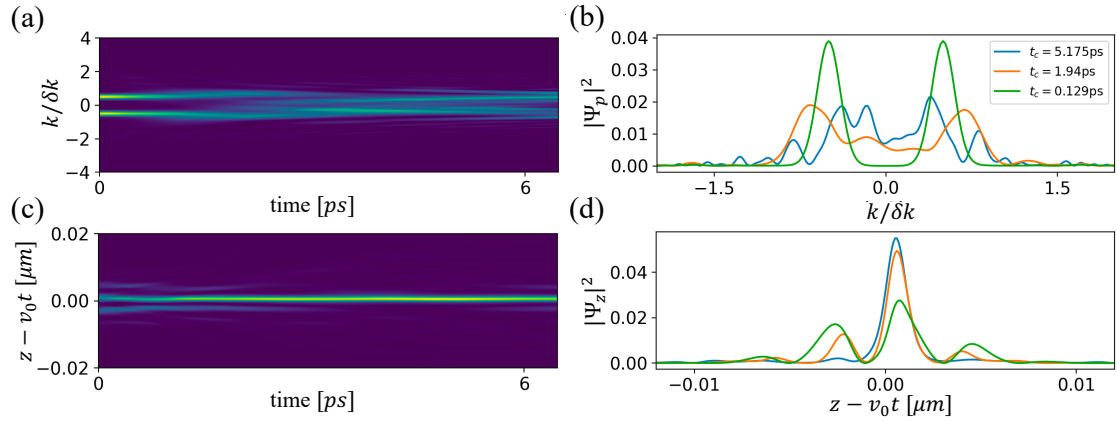

Fig. S1: TDSE simulation of the temporal profile of "half-electron (e/2)" by a twisted light wavefront, in both momentum (a,b) and spatial space (c,d). The right column displays snapshots corresponding to the left column, emphasizing the solitary characteristic of the light-induced "half-electron." Over time, the spatiotemporal profile becomes increasingly distinct, showcasing a pronounced dispersion-free property.

## 2. TDSE simulation algorithm

In order to assess our tight-binding-approximated result of a two-level electron in the Bragg regime, we have to solve the time-dependent Schrödinger equation (TDSE) directly

$$i\hbar \frac{\partial}{\partial t}\psi(z,t) = \widehat{H}\psi(z,t) = (\widehat{H}_0 + \widehat{H}_I)\psi(z,t), \tag{S2}$$

where the kinetic Hamiltonian of free electron is $\widehat{H}_0 = E_0 + v_0(\hat{p} - p_0) + \frac{(\hat{p}-p_0)^2}{2\gamma^3 m_e}$. We choose the initial kinetic energy $\varepsilon_0 = 100\ eV$, corresponding to the electron velocity $v_0 = \beta c$ with the relative speed $\beta = 0.02$ and the Lorentz factor $\gamma = 1.001$. We take a realistic nanograting, whose longitudinal vector potential is $A(z,t) = -\frac{E_z}{\omega_L}\sin(\omega_L t - k_z z + \theta(z,t))$, with electric field strength $E_z$, laser frequency $\omega_L$, wavevector $k_z = \frac{2\pi}{\Lambda}$ for a grating with a period $\Lambda$, and the phase $\theta$. Thus, the interaction Hamiltonian can be written as

$$\begin{aligned}\widehat{H}_I &= -\frac{e}{2\gamma m_e}[\hat{p}\cdot A(z,t) + A(z,t)\cdot\hat{p}] \\ &= -\frac{e}{\gamma m_e}A(z,t)\cdot\hat{p} - i\hbar k_z A_0 \cos(\omega_L t - k_z z + \theta(z,t)) \\ &\simeq -\frac{eA_0}{\gamma m}\sin(\omega_L t - k_z z + \theta(z,t))\cdot\hat{p}\end{aligned} \tag{S3}$$

where we disregard the second term under the approximation $p_0 \gg \hbar k_z$ (this will be demonstrated in the following), and $\theta(z,t) = -\frac{\pi}{2}\tanh(\frac{z-vt-z_0}{0.001})$.

We express the electron wavefunction as the slow-moving part with an initial phase:

$$\psi(z,t) = \chi(z,t)e^{-\frac{i(E_0 t - p_0 z)}{\hbar}} \tag{S4}$$

Thus, the left-hand side (LHS) of Eq. (S2) is

$$i\hbar\frac{\partial}{\partial t}\psi(z,t) = e^{-\frac{i(E_0 t - p_0 z)t}{\hbar}}\left(E_0 + i\hbar\frac{\partial}{\partial t}\right)\chi(z,t) \tag{S5}$$

Meanwhile, the right-hand side (RHS) becomes

$$\left[ e^{-\frac{i(E_0 t - p_0 z)t}{\hbar}} \left( E_0 + v_0((\hat{p} + p_0) - p_0) + \frac{((\hat{p} + p_0) - p_0)^2}{2\gamma^3 m_e} \right) \right.$$

$$-e^{-\frac{i(E_0 t - p_0 z)t}{\hbar}} \frac{eA_0}{\gamma m_e} \sin(\omega_L t - k_z z + \phi_0) \cdot (\hat{p} + p_0) \bigg] \chi(z,t) \tag{S6}$$

$$- \left[ e^{-\frac{i(E_0 t - p_0 z)t}{\hbar}} i\hbar k_z \frac{eA_0}{2\gamma m_e} \cos(\omega_L t - k_z z + \phi_0) \right] \chi(z,t)$$

$$\simeq e^{-\frac{i(E_0 t - p_0 z)t}{\hbar}} \left( E_0 + v_0 \hat{p} + \frac{\hat{p}^2}{2\gamma^3 m_e} - \frac{eA_0}{\gamma m_e} \sin(\omega_L t - k_z z + \theta(z,t)) \cdot (\hat{p} + p_0) \right) \chi(z,t)$$

here we omit the third term since $\frac{i\hbar k_z}{2} \cos(\omega_L t - k_z z + \phi_0) \ll p_0 \sin(\omega_L t - k_z z + \phi_0)$ under $p_0 \gg \hbar k_z$. Thus, we simplify the TDSE as

$$i\hbar \frac{\partial}{\partial t} \chi(z,t) = \left( \frac{\hat{p}^2}{2\gamma^3 m} + \left[ v_0 - \frac{eA_0}{\gamma m} \sin(\omega_L t - k_z z + \theta(z,t)) \right] \hat{p} \right) \chi(z,t)$$
$$- \frac{eA_0 p_0}{\gamma m} \sin(\omega_L t - k_z z + \theta(z,t)) \chi(z,t) \tag{S7}$$

Substitute the momentum operator $\hat{p} = -i\hbar \frac{\partial}{\partial z}$ and $p_0 = \gamma \beta mc$, $A_0 = \frac{E_0}{\omega_L}$ yields

$$i\hbar \frac{\partial}{\partial t} \chi(z,t) = \left( -i\hbar \left[ v_0 - \frac{eE_0}{\gamma m \omega_L} \sin(\omega_L t - k_z z + \theta(z,t)) \right] \frac{\partial}{\partial z} \right) \chi(z,t)$$
$$- \left[ \frac{\hbar^2}{2\gamma^3 m} \frac{\partial^2}{\partial z^2} + \frac{eE_0 \beta c}{\omega_L} \sin(\omega_L t - k_z z + \theta(z,t)) \right] \chi(z,t) \tag{S8}$$

The partial difference equation would be solved easier in the co-moving frame:

$$\zeta = z - v_0 t$$
$$t' = t$$

and we have

$$d\zeta = dz - v_0 dt \quad dz = d\zeta + v_0 dt'$$
$$dt' = dt \quad\quad\quad dt = dt'$$

which leads to the relations

$$\frac{\partial}{\partial z} = \frac{\partial}{\partial \zeta} \frac{\partial \zeta}{\partial z} + \frac{\partial}{\partial t'} \frac{\partial t'}{\partial z} = \frac{\partial}{\partial \zeta}$$
$$\frac{\partial}{\partial t} = \frac{\partial}{\partial \zeta} \frac{\partial \zeta}{\partial t} + \frac{\partial}{\partial t'} \frac{\partial t'}{\partial t} = \frac{\partial}{\partial t'} - v_0 \frac{\partial}{\partial \zeta}$$

In the co-moving frame, Eq. (S8) becomes

$$i\hbar\frac{\partial}{\partial t}\psi(z,t) = i\hbar\left(\frac{\partial}{\partial t'} - v_0\frac{\partial}{\partial \zeta}\right)\psi(\zeta,\tau_c)$$

$$= \left(-i\hbar\left[v_0 + \frac{eE_0}{\gamma m\omega_L}\sin(k_z\zeta - \theta(\zeta))\right]\frac{\partial}{\partial \zeta}\right)\chi(\zeta,\tau_c)$$

$$-\left[\frac{\hbar^2}{2\gamma^3 m}\frac{\partial^2}{\partial \zeta^2} - \frac{eE_0\beta c}{\omega_L}\sin(k_z\zeta - \theta(\zeta))\right]\chi(\zeta,\tau_c)$$

where $\theta(\zeta) = -\frac{\pi}{2}\tanh(\frac{\zeta-z_0}{0.001})$ and $\tau_c = ct'$. Divided by $\hbar c$ in both sides, we obtain

$$i\frac{\partial}{\partial \tau_c}\chi(\zeta,\tau_c) = -i\left[\frac{eE_0}{\gamma mc\omega_L}\sin(k_z\zeta - \theta(\zeta))\right]\frac{\partial}{\partial \zeta}\chi(\zeta,\tau_c)$$

$$-\left[\frac{\hbar}{2\gamma^3 mc}\frac{\partial^2}{\partial \zeta^2} - \frac{eE_0\beta}{\hbar\omega_L}\sin(k_z\zeta - \theta(\zeta))\right]\chi(\zeta,\tau_c) \quad (S9)$$

Notice that all the variables are calculated in the length unit of $\mu m$ and the time of $fs$. Assume that the electric field on the grating surface is $0.5 \times 10^7\,V/m$ and the incident laser $\lambda_L = 0.2\mu m$. Define

$$\alpha_1 = \frac{eE_0}{\gamma m_e c\omega_L} = 3.11 \times 10^{-7}, \quad \alpha_2 = \frac{\hbar}{2\gamma^3 m_e c} = 1.92 \times 10^{-7}\,\mu m,$$

$$\alpha_0 = \frac{eE_0\beta}{\hbar\omega_L} = 0.016\,\mu m^{-1}$$

The difference between the Bragg regime and Raman-Nath regime is determined by the factor

$$Q = \frac{\epsilon}{2|\kappa|} = \frac{\alpha_2 k_z^2}{\alpha_0} = \left(\frac{\hbar^2}{2m_e ec^3}\right)\frac{\omega_L^3}{\beta^3\gamma^3 E_0}$$

For our concern, we choose the parameters $Q = 29.3$ to obtain a two-level electron Rabi oscillation. As a result, the simplified equation is

$$i\frac{\partial}{\partial \tau_c}\chi(\zeta,\tau_c) = -i[\beta - \alpha_1\sin(k_z\zeta - \theta(\zeta))]\frac{\partial}{\partial \zeta}\chi(\zeta,\tau_c)$$

$$-\left[\alpha_2\frac{\partial^2}{\partial \zeta^2} + \alpha_0\sin(k_z\zeta - \theta(\zeta))\right]\chi(\zeta,\tau_c) \quad (S10)$$

Then we apply the discretization for the spatial parameter $\zeta$ by dividing the spatial domain into $N_\zeta$ parts, with the interval $\delta\zeta = (\zeta_{max} - \zeta_{min})/N_\zeta$, which leads to the following,

$$\frac{\partial^2}{\partial \zeta^2}\chi(\zeta,t) = \frac{1}{\delta\zeta^2}[\chi(\zeta+\delta\zeta,t) + \chi(\zeta-\delta\zeta,t) - 2\chi(\zeta,t)]$$

$$-i\beta\frac{\partial}{\partial\zeta}\chi(\zeta,t) = -\frac{i\beta}{2\delta\zeta}[\chi(\zeta+\delta\zeta,t) - \chi(\zeta-\delta\zeta,t)]$$

$$i\alpha_1\sin(k_z\zeta + \theta(\zeta))\frac{\partial}{\partial\zeta}\chi(\zeta,t) = -\frac{\alpha_1}{\delta\zeta}\cos(k_z\zeta + \theta(\zeta))\chi(\zeta,t)$$

$$+\frac{\alpha_1}{2\delta\zeta}\left[e^{i(k_z\zeta+\theta(\zeta))}\chi(\zeta+\delta\zeta,t) + e^{-i(k_z\zeta+\theta(\zeta))}\chi(\zeta-\delta\zeta,t)\right]$$

(S11)

Thus, the discretized Hamiltonian becomes

$$i\frac{\partial}{\partial\tau_c}\chi(\zeta,\tau_c) = \left[-\frac{\alpha_2}{\delta\zeta^2} - \frac{i\beta}{2\delta\zeta} + \frac{\alpha_1}{2\delta\zeta}e^{i(k_z\zeta+\theta(\zeta))}\right]\chi(\zeta+\delta\zeta,\tau_c)$$

$$+\left[-\frac{\alpha_2}{\delta\zeta^2} + \frac{i\beta}{2\delta\zeta} + \frac{\alpha_1}{2\delta\zeta}e^{-i(k_z\zeta+\theta(\zeta))}\right]\chi(\zeta-\delta\zeta,\tau_c) \quad (S12)$$

$$+\left[\frac{2\alpha_2}{\delta\zeta^2} - \frac{\alpha_1}{\delta\zeta}\cos(k_z\zeta+\theta(\zeta)) - \alpha_0\sin(k_z\zeta+\theta(\zeta))\right]\chi(\zeta,\tau_c)$$

In order to compute the time evolution of the slow-varying part $\chi(\zeta,\tau_c)$ for a given initial state $\chi(\zeta,\tau_0)$, we define the vector wavefunction with $N_\zeta$ components

$$v(\tau) = \left(\chi(\zeta_1,\tau), \chi(\zeta_2,\tau), \dots, \chi(\zeta_{N_\zeta},\tau)\right)^T \quad (S13)$$

Then, the differential equation (S12) can be simplified as

$$i\frac{\partial}{\partial\tau}v(\tau) = H(\tau)v(\tau) \quad (S14)$$

Applying the time discretization for the interval $\tau - \tau_0$ by equally dividing it into $N_t$ parts with the subinterval $\Delta\tau = (\tau - \tau_0)/N_t$ and we use an implicit Crank-Nicholson integrator to propagate the vector wavefunction from one time step to the next. The formal solution to Eq. (S14) can be expressed in terms of the time evolution operator

$$v(\tau + \Delta\tau) = U(\tau + \Delta\tau, \tau)v(\tau) \quad (S15)$$

where the time evolution operator can be expressed as

$$U(\tau + \Delta\tau, \tau) = (1 + i\,\Delta\tau H(\tau)/2)^{-1}(1 - i\,\Delta\tau H(\tau)/2) \quad (S16)$$

The solution of Eq. (S13) can give us the dynamics of the PINEM electron with an initial condition. The final wavefunction could be expressed as

$$v(\tau) = \prod_{n=1}^{N_t} U(\tau_0 + n\Delta\tau, \tau_0 + (n-1)\Delta\tau)\, v(\tau_0) \qquad (S17)$$